\documentclass[11pt,twocolumn]{article}

\usepackage{times}

\usepackage[labelfont=bf]{caption}

\usepackage[top=0.5in,bottom=0.75in,left=0.5in,right=0.5in]{geometry}

\usepackage{url}

\usepackage{subfig}

\usepackage{graphicx}

\usepackage{amsmath}

\graphicspath{{figures/}{plots/}{.}}

\usepackage[numbers]{natbib}
\usepackage[bookmarksopen,bookmarksnumbered,colorlinks=true,allcolors=blue]{hyperref}

\usepackage[compact]{titlesec}

\usepackage{helvet}
\usepackage{sectsty}
\allsectionsfont{\sffamily}

\begin{document}

%% ---------------------------------------------------------------------
%% Title, author(s), and affiliations
%% ---------------------------------------------------------------------

%% Paper title
\title{\sffamily Gatherplot: A Non-Overlapping Scatterplot}

%% Authors
\author{Deokgun Park,$^1$ Sung-Hee Kim,$^2$ and Niklas Elmqvist$^3$\\ 
\scriptsize $^1$University of Texas at Arlington, TX, USA; 
        $^2$Dong-eui University, Busan, Republic of Korea;
        $^3$University of Maryland, College Park, MD, USA}

\date{January 2023}

\maketitle

%% ---------------------------------------------------------------------
%% Abstract
%% ---------------------------------------------------------------------
\begin{abstract}
    Scatterplots are a common tool for exploring multidimensional datasets, especially in the form of scatterplot matrices (SPLOMs).
    However, scatterplots suffer from overplotting when categorical variables are mapped to one or two axes, or the same continuous variable is used for both axes.
    Previous methods such as histograms or violin plots use aggregation, which makes brushing and linking difficult.  
    To address this, we propose gatherplots, an extension of scatterplots to manage the overplotting problem.
    Gatherplots are a form of unit visualization, which avoid aggregation and maintain the identity of individual objects to ease visual perception.
    In gatherplots, every visual mark that maps to the same position coalesces to form a packed entity, thereby making it easier to see the overview of data groupings.
    The size and aspect ratio of marks can also be changed dynamically to make it easier to compare the composition of different groups.
    In the case of a categorical variable vs.\ a categorical variable, we propose a heuristic to decide bin sizes for optimal space usage.
    To validate our work, we conducted a crowdsourced user study that shows that gatherplots enable people to assess data distribution more quickly and more correctly than when using jittered scatterplots.
\end{abstract}

%% Keywords
\textbf{Keywords:} Multidimensional visualization; scatterplots; unit visualizations;  overplotting; scatterplot matrices (SPLOMs).

%% ---------------------------------------------------------------------
%% INTRODUCTION
%% ---------------------------------------------------------------------
% intro.tex

\section{Introduction}

\begin{figure}[htb]
    \centering
    \includegraphics[width=0.5\columnwidth]{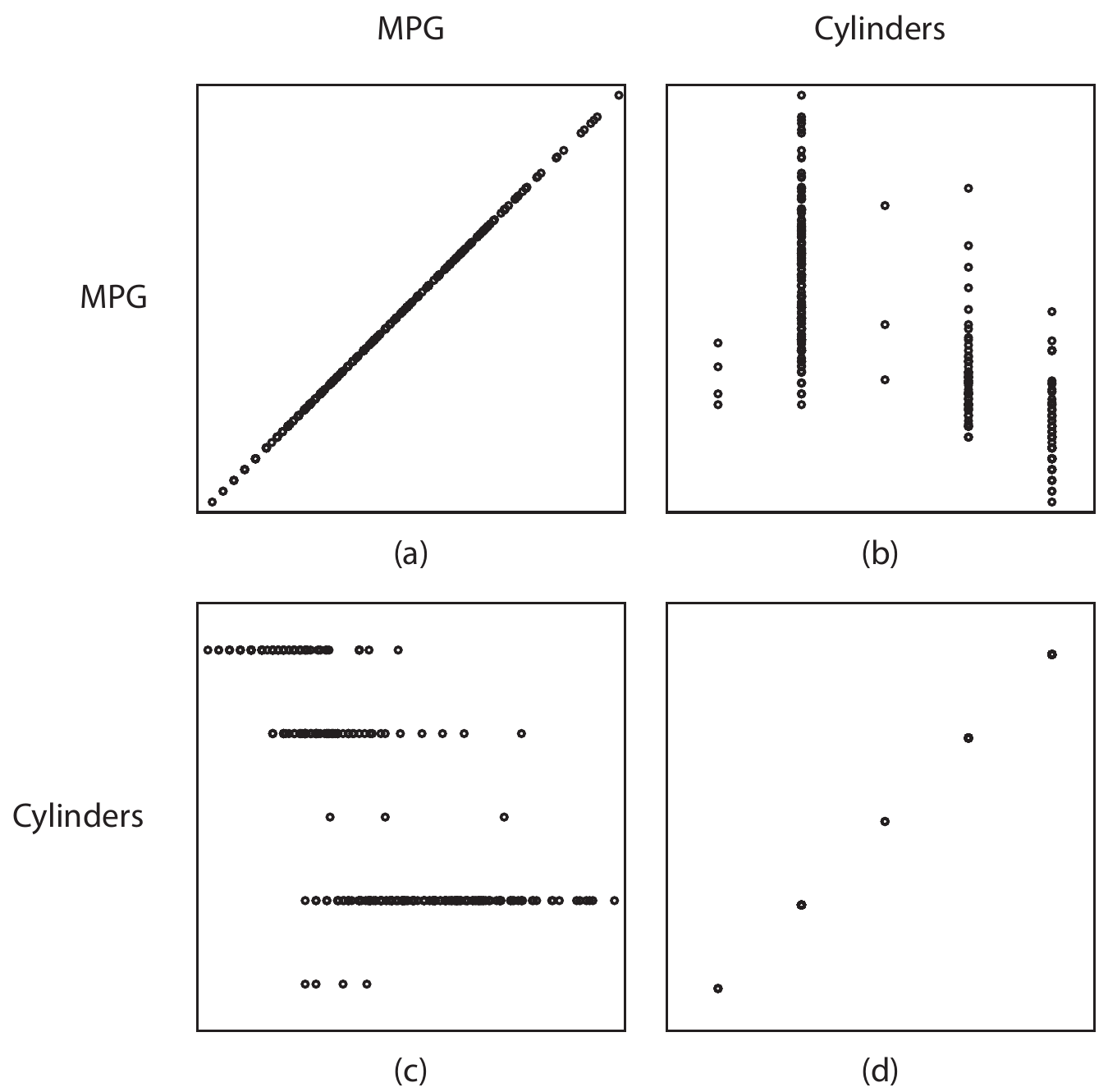}
    \caption{\textbf{Scatterplot matrix.}
    SPLOM visualizing a car dataset with one continuous variable \textit{MPG} and one categorical variable
    \textit{Cylinders} showing limitations of scatterplots when managing categorical variables.
    In (a), a scatterplot with the same variable for both axes results in a diagonal line.
    In (b) and (c), a scatterplot with a continuous vs.\ a categorical variable results in horizontal or vertical line patterns.
    In (d), a scatterplot with two categorical variables results in a dot pattern.}
    \label{fig:scatter-problems}
\end{figure}

Scatterplots---one of the most common types of statistical graphics~\cite{Cleveland1988, Elmqvist2008, Utts1996}---are often used to visualize two continuous variables using visual marks mapped to a two-dimensional Cartesian space, where the color, size, and shape of the marks can represent additional dimensions.
It can also be used for exploring multidimensional datasets in the form of scatterplot matrices (SPLOM), where all the possible
combinations of axes are presented in table form.
However, scatterplots are so-called \textit{overlapping} visualizations~\cite{Fekete2002} in that the visual marks representing individual data points may begin to overlap each other in screen space in situations when the marks are large, when there is insufficient screen space to fit all the data at the desired resolution, or simply when several data points share the same value.
In fact, realistic multidimensional datasets often contain categorical variables, such as nominal variables or discrete data dimensions with a small domain, which lead to many data points being mapped to the exact same screen position.
This kind of overlap is known as \textit{overplotting} (or \textit{overdrawing}) in visualization, and is problematic because it may lead to data points being entirely hidden by other points, which in turn may lead to the viewer making incorrect assessments of the data.
As can be seen in Figure~\ref{fig:scatter-problems}, there are three situations for mapping variables to axes in scatterplots when overplotting is inevitable:

\begin{itemize}
\item Plotting \textbf{categorical vs.\ continuous variables} gives rise to line patterns ((b) and (c) in Figure~\ref{fig:scatter-problems});
\item Plotting \textbf{categorical vs.\ categorical variables} gives rise to single dot patterns (Figure~\ref{fig:scatter-problems}(d)); and
\item Plotting the \textbf{same continuous variable on both axes} gives rise to diagonal line patterns (Figure~\ref{fig:scatter-problems}(a)).
\end{itemize}

Several approaches have been proposed to address this problem~\cite{Ellis2007}, the most prominent being transparency, jittering, and clustering techniques.
The first, changing transparency, does not so much address the problem as sidestep it by making the visual marks semi-transparent so that an accumulation of overlapping points are still visible. 
However, this does not scale for large datasets, and also causes blending issues if color is used to encode additional variables.
Jittering perturbs visual marks using a random displacement~\cite{Trutschl2003} so that no mark falls on the exact same screen location as any other mark, but this approach is still prone to overplotting for large data.
It also introduces uncertainty that is not aptly communicated by the scatterplot since marks will no longer be placed at their true location on the Cartesian space.
Other approaches still attempt to organize overlapping marks into visual groups that summarize their distribution, such as histograms, violin plots, and kernel density estimation (KDE) plots~\cite{Fua1999, Mayorga2013, Im2013}.
However, this comes at the cost of losing the identity of individual points, which can be problematic when filtering or searching; e.g, brushing data points is difficult in histograms~\cite{Im2013}.

\begin{figure*}[htb]
  \centering
  \resizebox{1.0\textwidth}{!}{\includegraphics{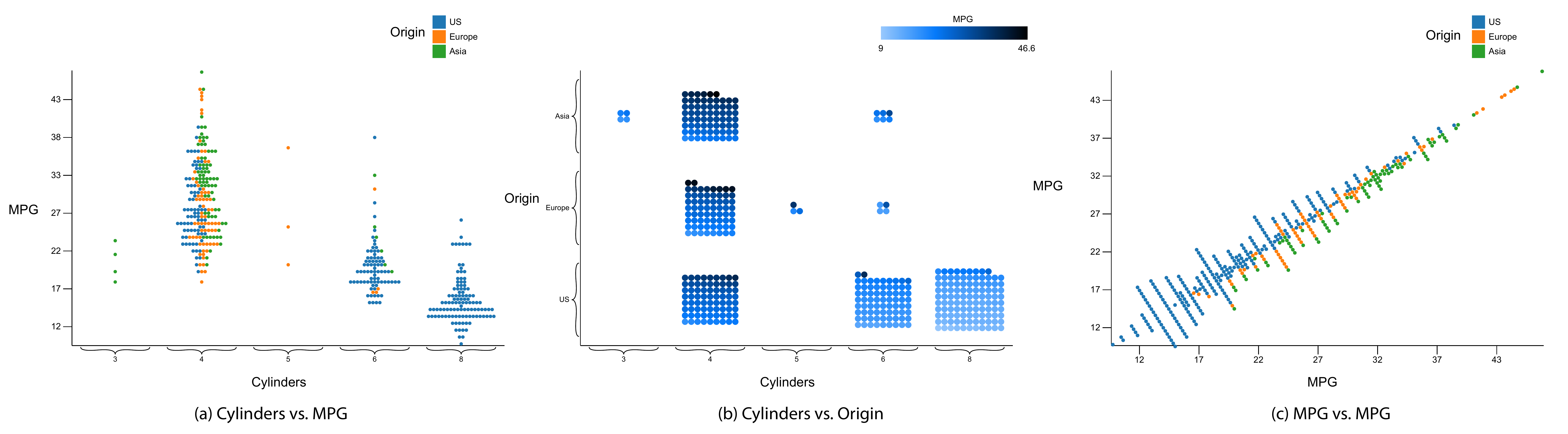}}
  \caption{\textbf{Cars dataset.}
  Gatherplots showing a dataset related to cars, yielding overplotting in normal scatterplots.
  The gatherplot in (a) shows Cylinders (categorical) vs.\ MPG (continuous), highlighting the overall distribution of MPG values of cars with different cylinders.
  The brackets on the X-axis are used to indicate that the interval within the brackets represent the same value in the data.
  The gatherplot in (b) shows \textsc{Cylinders} (categorical) vs.\ \textsc{Origin} (categorical), partitioning the graphical axes into intervals and packing points into groups for each interval. 
  In (c), both X-axis and Y-axis show the same continuous variable (MPG).
  All these cases would have caused overplotting for a scatterplot, resulting in dot-shaped or line-shaped point patterns where individual points cannot be identified.}
  \label{fig:teaser}
\end{figure*}

In this paper, we propose the concept of \textit{gathering} as an alternative to scattering and jittering, and then show how we can use this visual transformation to define a novel visualization technique called a \textit{gatherplot}.
The gatherplot is an instance of a recently recognized family of visualization techniques called \textit{unit visualizations}~\cite{Park2018} that maintain a strict mapping between every data item and its unique visual mark, improving the understandability over aggregated representations as well as enabling more natural interactions.
Gathering is a generalization of the linear mapping used by scatterplots, and works by first partitioning the graphical axis into segments based on the data dimension and then organizing points into \textit{packed groups} for each segment to avoid overplotting.
This means that the gather operation relaxes the continuous spatial mapping commonly used for a graphical axis; instead, each discrete segment occupies a certain interval of screen space that maps to the same data value.
This is communicated using graphical brackets on the axis that shows the value or interval for each segment (Figure~\ref{fig:teaser}(b)).
Beyond the gatherplot technique, we also show how the gather transformation can be embedded into a Magic Lens~\cite{Bier1993} for use in a normal scatterplot; we call this technique a ``Gather Lens.''

The contributions of our paper are the following: (1) the concept of the gather visual transformation as a generalization of linear visual mappings; (2) the gatherplot technique, an application of the gather operation to scatterplots to mitigate overplotting; (3) the GatherLens interaction technique where gathering is applied on a local level; and (4) results from a crowdsourced graphical perception study on the effectiveness of gatherplots.

Next in this paper, we review the literature on statistical graphics and overplotting.
We then present the gather operation and use it to define gatherplots.
We describe our implementation, followed by our crowdsourced evaluation.
We finally describe the GatherLens technique and close with conclusions, and our future plans.

%% ---------------------------------------------------------------------
%% BACKGROUND
%% ---------------------------------------------------------------------
\section{Background}

Our goal with gatherplots is to generalize scatterplots to a representation that maintains its simplicity and familiarity while eliminating overplotting.
With this in mind, below we review prior art on mitigating overplotting using appearance or distortion.
We also discuss visualization techniques specifically designed for categorical variables.

\subsection{Characterizing Overplotting}

While there are many ways to categorize visualization, Fekete and Plaisant~\cite{Fekete2002} introduced a classification particularly useful for our purposes that splits techniques into two types:

\begin{itemize}

\item\textbf{Overlapping visualizations:} No layout restrictions on visual marks is enforced, leading to overplotting.
Examples: Scatterplots, node-link diagrams, parallel coordinate plots, etc.

\item\textbf{Space-filling visualizations:} Layouts designed to optimally fill the available space to avoid overlap.
Examples: Treemaps, matrices, geographic maps.

\end{itemize}

Fekete and Plaisant~\cite{Fekete2002} investigated the overplotting phenomenon for a 2D scatterplot, and found that it has a significant impact as datasets grow.
The problem stems from the fact that even with two continuous variables that do not share any coordinate pairs, the size ratio between the visual marks and the display remains mostly constant.
Furthermore, most datasets are not uniformly distributed.
This all means that overplotting is inevitable for realistic datasets.

Partial or complete overplotting generally leads to visual clutter.
Ellis and Dix~\cite{Ellis2007} survey the literature and derive a general approach to reduce clutter.
According to their treatment, there are three ways to reduce clutter in a visualization: by changing the visual appearance, by distorting visual space, or by presenting data over time.
Some trivial but impractical mechanisms they list include decreasing mark size, increasing display space, or animating the data.
Below we review more practical approaches based on appearance and distortion.

\subsection{Appearance-based Methods}

Practical appearance-based approaches to mitigate overplotting include transparency, sampling, kernel density estimation (KDE), and aggregation.
Transparency changes the opacity of the visual marks, and has been shown to convey overlap for up to five occurrences~\cite{Zhai1996}.
However, there is still an upper limit for how much overlap is perceptible to the user, and the blending caused by overlapping marks of different colors makes identifying colors difficult.

\textit{Sampling} uses stochastic methods to statistically reduce the data size to visualize~\cite{Dix2002}.
This may reduce the amount of overplotting, but since the sampling is random, it cannot be reliably eliminated.
Furthermore, one of the core strengths of a scatterplot is its ability to show outliers effectively, whereas sampling will likely eliminate all outliers (due to the intrinsic nature of an outlier).

Aggregation methods can also mitigate overplotting.
KDE~\cite{Silverman1986} and other binned aggregations~\cite{Elmqvist2010, Fua1999, Mayorga2013, Im2013} replace a cluster of marks with a single entity that has a distinct visual representation. 
Similarly, splatterplots~\cite{Mayorga2013} combine individual marks with aggregated entities, using marks to show outliers and aggregated entities to show the general trends.
While aggregation techniques are effective against overplotting for continuous variables, they fare poorly for categorical ones.
Therefore, the generalized plot matrices (GPLOMs)~\cite{Im2013} were proposed to solve this particular problem by adopting non-homogeneous plots into a matrix.
The technique uses a histogram for categorical vs.\ continuous variables, and a treemap for categorical vs.\ categorical variables.
While effective in providing overview, aggregated techniques sacrifice some compatibility with scatterplots since they no longer maintain object identity, meaning that each visual mark no longer represents a single data point.  

\subsection{Distortion-based Methods}

Distortion-based techniques avoid overplotting by changing the spatial mapping of the space and have the advantage to keep the identity of individual data points.
The canonical distortion technique is \textit{jittering}, where a random displacement is used to subtly modify the exact screen space position of a data point.
This has the effect of spreading data points apart so that they are easier to distinguish.
However, na{\"i}ve jittering mechanisms apply the displacement indiscriminately to all data points, regardless of whether they are overlapping or not.
This has the drawback of distorting points away from their true location on the visual canvas, and still does not completely eliminate overplotting.

Bezerianos et al.~\cite{Bezerianos2010} use a more structured approach to displacement, where overlapping marks are organized onto the perimeter of a circle.
The circle is grown to a radius so that all marks fit, which means that its size is also an indication of the number of grouped points.
However, this mechanism still introduces uncertainty in the spatial mapping, and it is also not clear how well it scales for very dense data, as this can lead to a circle of arbitrarily large size.
Nevertheless, the approach is a good example of how deterministic displacement can be used to great effect for eliminating overplotting.

Trutschl et al.~\cite{Trutschl2003} propose a deterministic displacement (``smart jittering'') that adds meaning to the jittered position based on clustering results.
This makes it easier to understand the resulting spatial display.

\subsection{Data-aware Methods}

The most advanced and effective overplotting mitigations are data-aware, in that they determine instances of overplotting in a chart.
As a case in point, very recent work by Chen et al.~\cite{Chen2018} use animation to cycle the depth value of overlapping points in a scatterplot over time to ensure that every point is shown on top at some point in the rotation.
This means that overplotting is alleviated by the notion of ``guaranteed visibility over time''~\cite{Munzner2003} presented by Munzner et al.

Shneiderman et al.~\cite{Shneiderman2000} propose a data-aware structured displacement approach called \textit{hieraxes}, which combines hierarchical browsing with two-dimensional scatterplots.
In hieraxes, a two-dimensional visual space is subdivided into rectangular segments for different categories in the data, and points are then coalesced into stacked groups inside the different segments.
This work inspired gatherplots, which refines the layout and design of hieraxes further.

Microsoft's SandDance~\cite{Sanddance} use atomic visual marks as the building block of a highly interactive and visual interface built on smooth transitions between different spatial mappings.
Drawing on an older experimental tool called Pivot from Microsoft Live Labs, SandDance now exists as a custom visual in the Microsoft Power BI tool.\footnote{\url{https://www.microsoft.com/en-us/garage/profiles/sanddance/}}

Finally, Keim et al.~\cite{DBLP:journals/ivs/KeimHDJB10} propose \textit{generalized scatterplots} that use a data-aware combination of overlapping and distortion to avoid overplotting in a scatterplot display.
By balancing the overlapping and distortion, the user can achieve a display that conforms to their prior familiarity with scatterplots while retaining minimal occlusion and appropriate distortion of data points.
However, in contrast, our gatherplot approach requires no such balancing; of course, as a result, it is also less visually scalable.

Hieraxes~\cite{Shneiderman1996}, SandDance~\cite{Sanddance}, and gatherplots (that we present in this paper) are all examples of a recently recognized family of visualizations called \textit{unit visualizations}~\cite{Park2018} where the relation between data items and their mark is explicitly maintained. 
This identity property between data and display is exemplified in visualizations such as unit charts, dotplots, and scatterplots.
It can be contrasted with \textit{aggregated visualizations} that combine multiple data items intop a single visual mark, such as barcharts, piecharts, and histograms.
Our gatherplots technique shows how a unit visualization can be designed, evaluated, and even deployed in an interactive technique (GatherLens) from the ground up based on unit visualization principles.

\subsection{Visualizing Categorical Variables}

While we have already ascertained that scatterplots are not optimal for categorical variables, there exists a multitude of visualization
techniques that have been specifically designed for such data.~\cite{Bederson2002, Hofmann2000, Kosara2006}
Simplest among them are histograms, which visualize the item count for each categorical value~\cite{Stevens1946}.
Boxplots and violin plots show the distribution of continuous variables over categorical variables~\cite{Wickham2011}.
While hieraxes, histograms, and treemaps are effective in dealing with categorical variables, it is difficult to extend these to continuous vs.\ categorical variables.
One way is to apply binning to continuous variables to create groups of values.
However, the optimal number of bins depends on statistical characteristics of the data and the required task.
Dot plots by Wilkinson~\cite{Wilkinson1999} renders continuous univariate variables without overplotting by stacking nodes within dot
size.
Dang et al.~\cite{Dang2010} extended this to scatterplots by stacking nodes whose values are similar in 3D visual space.
These pioneering works provide the theoretical background for the determination of optimal bin size for gatherplots.    

Another method for visualizing categorical data that is of practical interest is for making inferences based on statistical and probabilistic data.
Cosmides and Toody~\cite{Cosmides1996} used frequency grids as discrete countable objects, and Micallef et al.~\cite{Micallef2012} extend this with six different area-proportional representations of categorical data organized into different classes.
Huron et al.~\cite{Huron2013} suggested using sedimentation as metaphor where individual objects coming from a data stream gradually transform into aggregated areas, or strata.

\begin{figure*}[htb]
    \centering
    \includegraphics[width=\textwidth]{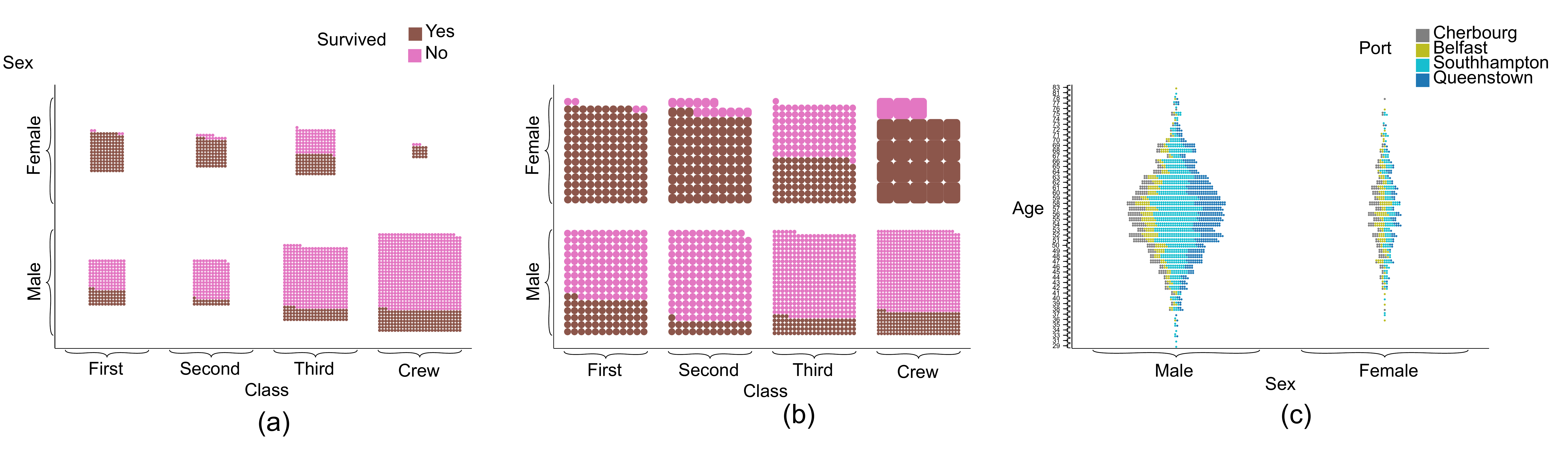}
    \caption{\textbf{Main layout modes for gatherplots.}
    (a) absolute mode with constant aspect ratio, which maintains the aspect ratio; (b) normalized mode of (a).
    The rate of male survivors in each passenger class is not easy to compare.
    Figure (c) shows the streamgraph mode, where each cluster maintains the number of element in the shorter edge, making it easier to see the distribution of the subgroups along the Y axis.}
  \label{fig:aspectRatio}
\end{figure*}

%% ---------------------------------------------------------------------
%% THEORETICAL BASIS
%% ---------------------------------------------------------------------
%% ---------------------------------------------------------------------
%% THEORETICAL BASIS
%% ---------------------------------------------------------------------
\section{The Gather Transformation}
\label{sec:gather}

Position along a common scale is the most salient of all visual
variables~\cite{Bertin1983, Cleveland1985}, and so mapping a data
dimension to positions on a graphical axis is a standard operation in
data visualization.
We call this mapping a \textit{visual transformation}.
However, most statistical treatments of data, such as Stevens'
classical theory on the scale of measurements~\cite{Stevens1946}, do
not take the physical properties of display space into account.
This is our purpose in the following section.

\subsection{Problem Definition}

Let $V = <f, s>$ be a visual transformation that consists of a
transformation function $f$ and a mark size $s$ (pixels).
Furthermore, assume that $f$ transforms a data point $p_d \in D$ from
a data dimension $D$ to a coordinate on a graphical axis $p_c \in C$
by $f(p_d) = p_c$.
Given a dataset $D_i \subseteq D$, we say that a particular visual
transformation $V_j$ exhibits \textit{overlap} if

$$
\exists p_x, p_y \in D_i \wedge x \neq y : |f_j(p_x) - f_j(p_y)| < s_j.
$$

In other words, overlap occurs for a particular dataset and visual
transformation if there exists at least one case where the visual
marks of two separate data points in the dataset fall within the same
interval on the graphical axis.
The \textit{overlap index} of a dataset and visual transformation is
defined as the number of unique pairs of points that overlap.
For a one-dimensional visualization, only a single transformation is
used and the visualization and dataset is said to exhibit
\textit{overplotting} iff it exhibits overlap.
For a two-dimensional visualization, however, the visualization and
dataset will only exhibit overplotting iff there is overlap in
\textbf{both} visual transformations and data dimensions.
Analogously, the \textit{overplotting index} is the unique number of
overplotting incidences for that particular visual transformation and
dataset.

This has two practical implications: (1) even a dataset that consists
only of nominal variables may \textbf{not} exhibit overplotting if
there is only at most one instance of each nominal value, and (2) a
dataset consisting of continuous values may \textbf{still} exhibit
overplotting if any two points in the dataset are close enough that
they get mapped to within the size of the visual marks on the screen.
The corollary is basically that overplotting is a function of
\textbf{both} visualization technique and dataset.

\subsection{Definition: The Gather Transformation}

We build on the previous idea of structured
displacement~\cite{Bezerianos2010, Shneiderman2000} by proposing a
novel visual transformation function called a \textit{gather}
transformation $f_{gather}$ that non-linearly segments the graphical
axis $C$ and organizes data points in each segment to eliminate
overplotting.

The gather transformation $V_{gather} = <f_{gather}, s_{gather}>$
consists of a transformation function $f_{gather}$ that maps data
points $p_d \in D$ to coordinates $p_c \in C$, and a visual mark
sizing function (instead of a scalar) $s_{gather}$ that yields a
visual mark size given the same data point.
The gather transformation function is special in that it eliminates
overplotting by subdividing the graphical axis $C$ into $n$ contiguous
segments $C = \{ C_1, C_2, \ldots, C_n \}$, where $n$ is the size of
the domain of the gather transformation function, i.e., the number of
unique elements in the data dimension $D$.
When mapping a data point $p_d$ to the graphical axis, $f_{gather}$ will
return an arbitrary graphical coordinate $p_c \in C_i$ for whatever
coordinate segment $C_i$ that $p_d$ belongs to.

Practically speaking, coordinates $p_c \in C_i$ will be chosen to
efficiently pack visual marks into the available display space without
causing overplotting (i.e., using a regular spacing of size
$s_{gather}$).
Several different methods exist for adapting the gather transformation
to the dataset $D$.
One approach is to keep the segments $C_1, \ldots, C_n$ of equal size
and find a constant visual mark size $s_{gather}(p_d) = s_{max}$ that
ensures that all points fit within the most dense segment.
The constant mark size makes visual comparison straightforward.
Another approach is to adapt segment size to the density of the data
while still keeping the mark size constant.
This will minimize empty space in the visual transformation and allows
for maximizing mark size.
A third approach is to vary mark size proportionally to the number of
points in a segment.
This will make comparison of the absolute number of points in each
segment difficult, but may facilitate relative comparisons if marks
are distinguished in some other way (e.g., using color).

For data dimensions $D$ that have a very large number of unique
values, it often makes sense to first quantize the data using a
function $p_q = Q(p_d)$ so that the number of elements $n$ is kept
manageable (on the order of 10 or less for most visualizations).
For example, a data dimension representing a person's age might
heuristically be quantized into ranges of 10 years: 0-9 years, 10-19
years, 20-29 years, and so on.

In a gather transformation, the coordinate axis has been partitioned
into segments, where the order of segments on the axis depends on the
data.
For nominal data, the segments can be reordered freely, both by the
algorithm and by the user.
For ordinal or quantized data, the order is given by the data
relation.
Furthermore, it often makes sense to be able to order points inside
each segment $C_i$ using the gathering transformation function
$f_{gather}$, for example using a second data dimension (possibly
visualized using color) to group related items together.

Appropriate visual representations of data where the gather
transformation has been applied are also important.
The \textit{stacked entities} of gathered points---one per coordinate
segment $C_i$---should typically maintain object identity, so that
each constituent point and their size is discernible as a discrete
visual mark.
Similarly, a visual representation of the segmented graphical axis
should externalize the segments as labeled intervals instead of
labeled major and minor ticks; this will also communicate the
discontinuous nature of the axis itself to the viewer.
 
\subsection{Using the Gather Transformation}

To give an example in one-dimensional space, parallel coordinate
plots~\cite{Inselberg1985} use multiple graphical axes, one per
dimension $D_i$, and organize them in parallel while rendering data
points as polylines connecting data values on one axis to adjacent
ones.
However, traditional parallel coordinate plots merely use a scatter
transformation on each graphical axis, which makes the technique prone
to overplotting.
Multiple authors have studied ways of mitigating this problem, for
example by reorganizing the position of nominal
values~\cite{Rosario2004}, using transparency, applying jitter, or by
clustering the data~\cite{Fua1999}.

However, an alternative approach is to use the gather representation
for each graphical axis to minimize overplotting.
This will cause each axis to be segmented into intervals, and we can
then resize segments according to the number of items falling into
each segment so that segments with many data points become
proportionally larger than those with fewer points.
Finally, if the data dimensions represent nominal data, it may make
sense to use a global segment ordering function so that there is a
minimum of lateral movement for the majority of points as they
connect to adjacent axes.
This will also minimize line crossings between the parallel axes.
This particular visualization technique---a parallel coordinate plot
with the gather transformation applied to each graphical axis---is
essentially equivalent to \textit{parallel sets}~\cite{Kosara2006}.

In fact, by applying our generalized gather transformation to the
axis, we are actually proposing a new type of stacked visualization
where each entity is still represented by lines.
In a sense, this technique combines parallel coordinates and parallel
sets because the grouped lines maintain the illusion of a single
entity for an axis with nominal categorical values (similar to
parallel sets), yet integrates directly with a parallel coordinate
axis with continuous values.
The main difference is that the new parallel coordinate/set variation
allows each axis to be either categorical or continuous, meaning that
one axis can represent the gender and the next can represent the
height of person.

%% ---------------------------------------------------------------------
%% GATHERPLOTS
%% ---------------------------------------------------------------------
\section{Gatherplots: A 2D Gathering Representation}

Applying gathering to two perpendicular axes defining a Cartesian space results in a \textit{gatherplot}: a 2D distortion-based extension of scatterplots that gathers data points into \textit{stacked groups}, thereby eliminating overplotting without losing the identity of individual data points.
Compared to jittering, which relies on random permutation, gathering organizes visual marks according to visual features, so that the resulting group of objects forms a meta-object.
According to Haroz and Whitney~\cite{Haroz2012}, grouping marks by feature helps in performing perceptual tasks such as finding outliers, counting items, seeing trends, and so on.
The technique is particularly designed for visualizing categorical variables.
Below we discuss the open design parameters for the technique, including layout, aspect ratio, and item shapes.

\subsection{Layout} 
 
Gatherplots eliminate overplotting by gathering marks with similar visual properties into \textit{stacked groups}.
This is inspired by previous works such as hieraxes~\cite{Shneiderman2000} or frequency grids~\cite{Micallef2012, Cosmides1996}.
However, there are many design possibilities for organizing the visual representation depending on the context, especially on the size distribution of each groups, the aspect ratio of assigned space, and the task at hand.  
As a result, we derive the following three layout modes (see Figure~\ref{fig:aspectRatio}):
 
\begin{itemize}

\item\textbf{Absolute mode:} Here stacked groups are sized to follow the
  aspect-ratio of the assigned region.
  The size of the items are determined by the maximum length dots
  which can fill the assigned region without overlapping.
  This means with the same assigned space, the groups with the maximum
  number of members determines the overall size of the nodes
  (Figure~\ref{fig:aspectRatio}(a)). 

\item\textbf{Normalized mode:} In this mode, the mark size and aspect
  ratio is adapted so that every stacked group has equal dimensions.
  This makes it easier to investigate ratios when the user is
  interested in the relative distributions of subgroups rather than
  the absolute number of members.
  Items also change their shape from a circle (absolute mode) to a
  rounded rectangle (Figure~\ref{fig:aspectRatio}(b)).  
  
  Normalized mode is useful for two specific tasks:

  \begin{itemize}
    
  \item Finding the ratio of the subgroups in a group
    (Figure~\ref{fig:aspectRatio}).
    Because groups of different size are normalized to the same
    geometric area, any comparison results in a relative comparison,
    which can aid statistical Bayesian reasoning~\cite{Micallef2012}.
  
  \item Finding the distribution of outliers.
    When there are many items on the screen for absolute mode, all
    marks must be reduced in size.
    This can make outliers hard to locate.
    When normalized mode is used, the outliers are expanded to fill
    the assigned space, making them easier to see.  
  \end{itemize}

\item\textbf{Streamgraph mode:} Here stacked groups are reorganized
  so that they maintain the same number of elements in their shorter
  edge.
  This mode is used for regions where the ratio of width and height
  are drastically different (in our prototype implementation, we use a
  heuristic threshold aspect ratio value of 3 for activating this
  mode).
  This means there are usually many times more groups in the axis in
  parallel with shorter edges.
  A good example is for visualizing the population distribution with
  regards to gender and age; the resulting gatherplot approaches
  ThemeRiver~\cite{Havre2000} as the number of entities increases
  (Figure~\ref{fig:aspectRatio}(c)).
  
\end{itemize}
 
 The choice between absolute and streamgraph mode happens
 automatically based on the aspect ratio of assigned space and without
 the need for user intervention.
 Therefore, only a simple interaction is required to toggle between
 absolute and normalized mode.
 However, we do expose a setting to manually toggle between these
 modes as well.

\subsection{Managing Continuous Variables} 

To use gatherplots for continuous variables, we apply binning to
partition the variable into discrete intervals.
The resulting visualization resembles dots plots by
Wilkinson~\cite{Wilkinson1999}, where bin size is equal to dot
size.
The size of individual bins is important for binning because it
determines the spatial accuracy and legibility of the visualization.  

\begin{figure}[htb]
  \centering
  \includegraphics[width=\columnwidth]{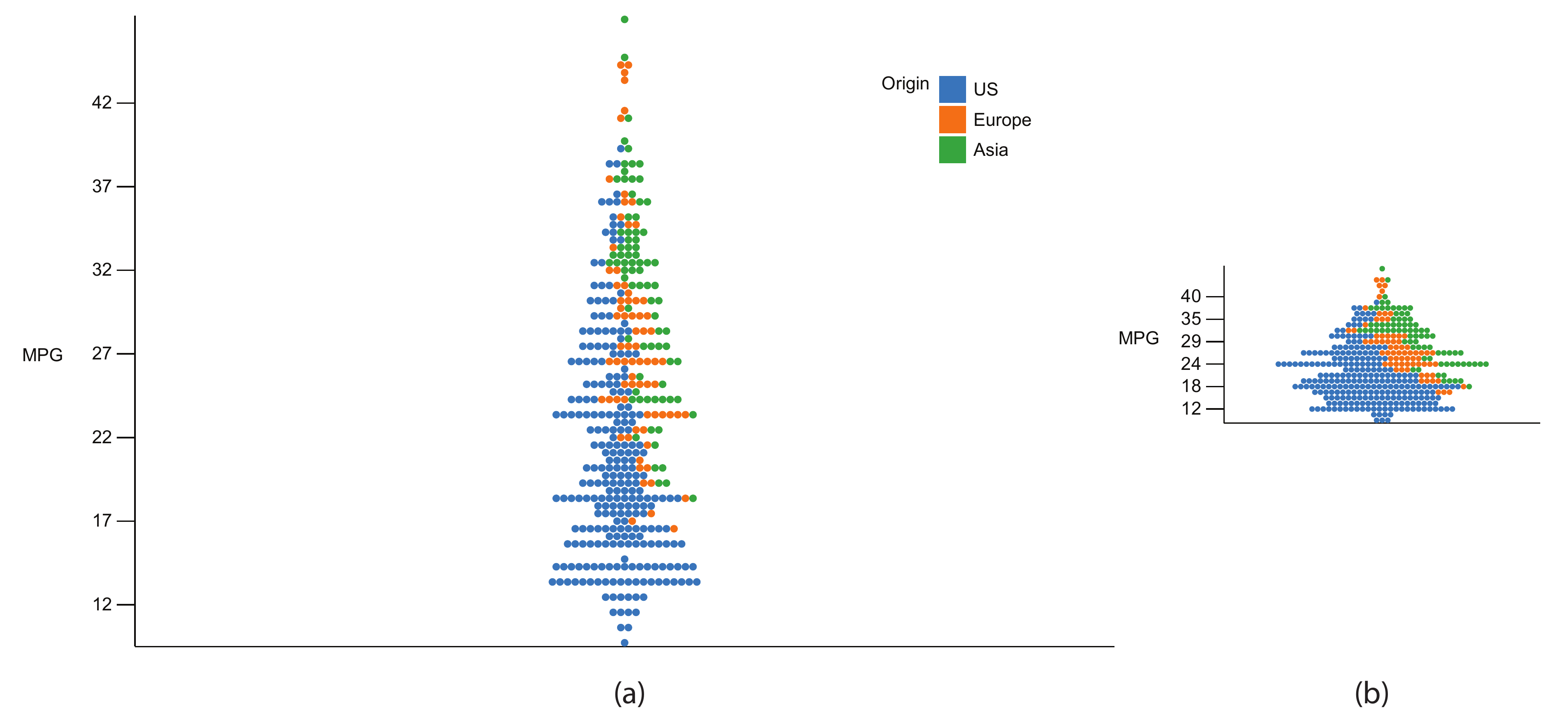}
  \caption{\textbf{Choosing optimal bin size based on available display space.}
    In (a), there is enough space so that the dot size can be
    maximized, improving spatial accuracy.
    In comparison, in (b) the assigned space is small, so the dot size
    is determined so that the most crowded bin interval will fit
    within the assigned space.
    This results in two different overviews even though the two plots
    have identical aspect ratio.}
  \label{fig:optimalbin}
\end{figure}

Wilkinson proposed $.25n^{-1/2}$ as the optimal dot size for dot
plots.
This creates reasonable dot plots for fixed aspect ratio of 5 to 1,
which is common in statistical charts assuming normal distribution of
nodes.   
However, gatherplot requires two different assumptions: First, the
aspect ratio varies according to the space given to the categorical
variables.  
Second, the dot size or bin size is determined by the global maximum
in the dataset, which may not be in the same cluster.
Furthermore, because bin size is the same as dot size, selecting bin
size can be thought of as a trade-off between accuracy and legibility.
Using very small bin size and dot size increases the spatial accuracy,
but results in poor legibility, and vice versa.
Balancing accuracy vs.\ legibility is common in visualization for
large datasets; for example, splatterplots limit the information shown
to users based on the available visual space~\cite{Mayorga2013}.
Similarly, gatherplots choose bin size based on spatial accuracy and
legibility.
When the visual space is small, we use a comparably large bin size to
increase dot size, thus resulting in poor spatial accuracy and high
legibility; for larger space allocations, the bins can be made smaller
to increase accuracy without loss of legibility.
This is shown in Figure~\ref{fig:optimalbin}. 

\begin{figure*}[htb]
  \centering
  \includegraphics[width=\linewidth]{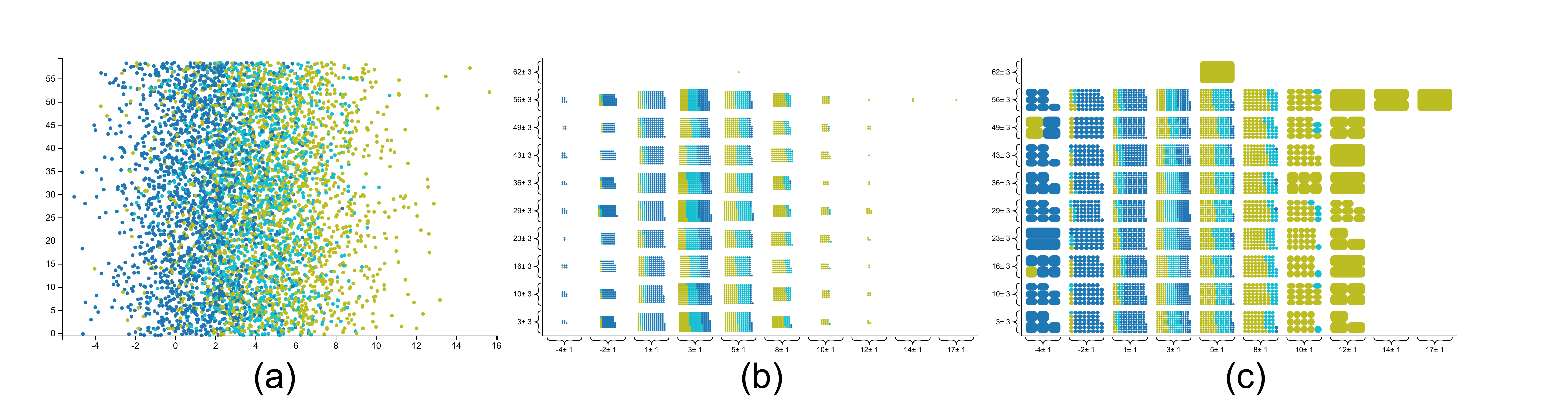}
  \caption{\textbf{Using gatherplots to manage overplotting.}
    (a) shows a scatterplot with 5,000 random numbers with severe
    overplotting in the center area.
    In (b), gathering is applied to create a more organized view.
    However, the gathering resizes the items so small that it becomes
    difficult to detect outliers.
    (c) shows normalized mode, where the outliers are enlarged.
    This makes identifying the distribution of sparse regions easier.}
 \label{fig:contcont}
\end{figure*}

Figure~\ref{fig:contcont} shows how gatherplots handle the situation when continuous variables are assigned to \textit{both} axes, causing both to be binned.
The plot is using normalized mode with two random variables.
The normalized mode makes it easier to identify the outliers and the distribution of outliers.
Furthermore, the case of scatterplots with the same continuous variables on both axes can be treated as a special case of continuous vs.\ categorical variables.
Here, the gatherplot is rotated to maintain integrity with scatterplots (Figure~\ref{fig:teaser} (c)). 

One limitation of gatherplots is that the technique requires binning to manage a continuous variable, yet binning creates arbitrary
boundaries that can be misleading.
However, using both gatherplots and scatterplots in different views
makes this problem less severe because the analyst can simply choose
the visual representation most suited for a particular task.

\subsection{Undefined Axis Mapping}

\begin{figure}[htb]
  \centering
  \includegraphics[width=\linewidth]{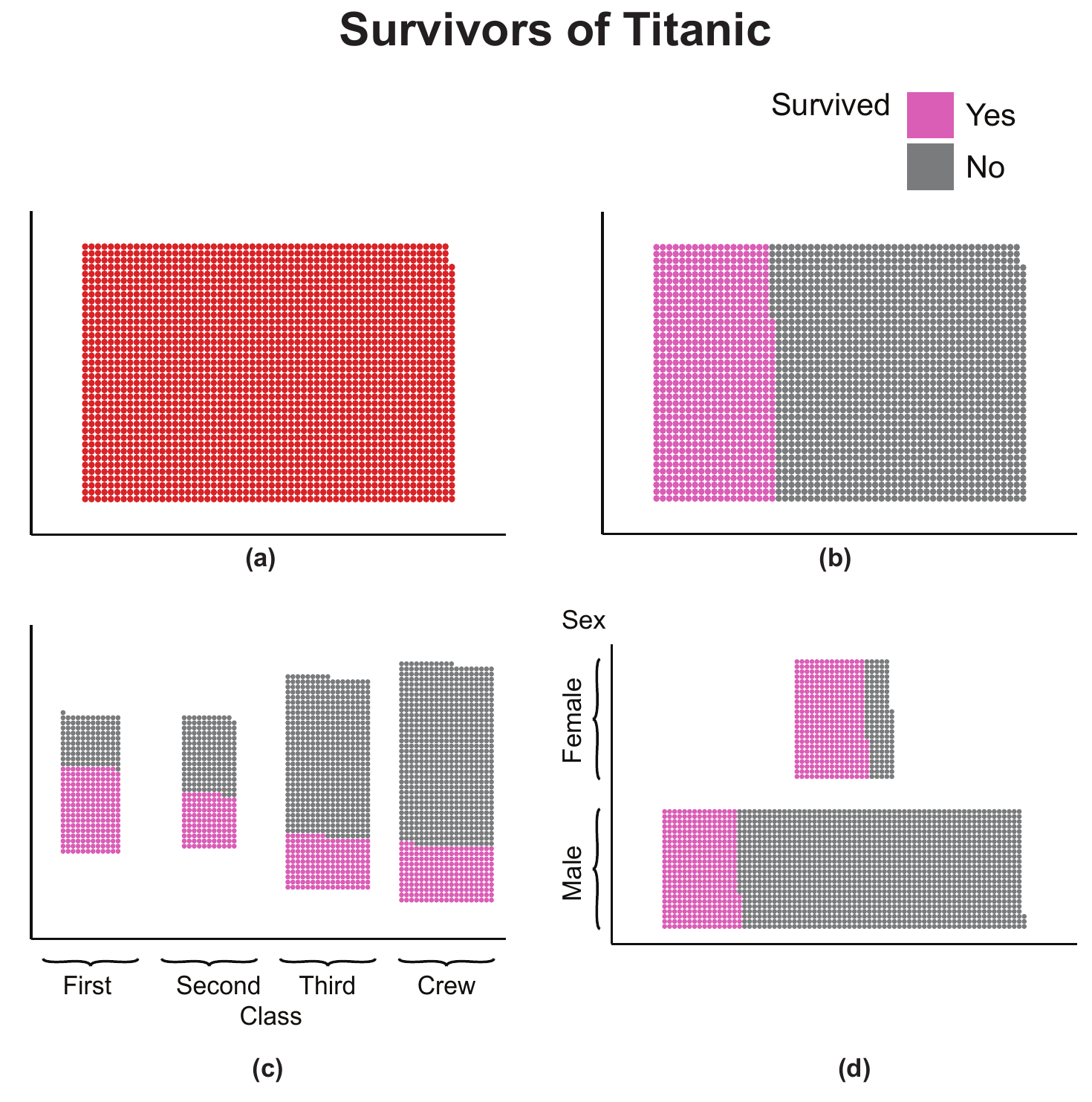}
  \caption{\textbf{Gatherplot showing survivors of the \textit{Titanic}.}
    (a) All people on board.
    Note that the X and Y axis are not defined.
    (b) Color coding for survivors.
    (c) Distribution of survivors over class variables.
    Here the Y axis is undefined.
    (d) Distribution of survivors over the gender.
    Here the X axis is undefined.}
  \label{fig:matrix}
\end{figure}

Scatterplots have traditionally been used to view correlations between two variables.
However, for a multidimensional exploration, one subtle difficulty is when the user wants to see only the effect of a single variable.
In gatherplots, the logical extension of an undefined axis is the aggregation of all nodes in a single group along that axis.
Figure~\ref{fig:matrix} shows an example of this using a dataset on survivors of the Titanic.

\subsection{Visual Design}

Gatherplots build on the same visual language as scatterplots.
However, some aspects are different; below we discuss our design choices for visual mark shape as well as tick marks.

\subsubsection{Visual Marks}

Scatterplots typically use a small circle or dot as a visual
representation for items, but many variations exist that use glyph
shapes to convey multidimensional variables~\cite{McDonnel2009,
  Tufte1983, Carr1983, Cleveland1988, Chernoff1973}.
However, in normalized mode, sometimes the aspect ratio of visual
marks changes according to the aspect ratio of the space assigned to
that value.
Also, as gathering changes the size of marks to fit in one cluster,
sometimes the marks size becomes too small or too large compared to
other marks.
This results in several unique design considerations for item shapes.

Based on our experience with several alternate designs, we recommend using a rectangle with constant rounded edge without using stroke lines.
Using constant rounded edge allows the nodes to be circular when the mark is small, as in Figure~\ref{fig:aspectRatio}(b), and a rectangle to show the degree of stretching, as shown in Figure~\ref{fig:aspectRatio}(b).
As a disadvantage, it does yield less differentiable individual elements.
As for avoiding stroke lines, such lines become dominant when nodes shrink below a certain size.
Eliminating them entirely curtails this problem.
 
\subsubsection{Interval Tick Marks} 
 
Because we are representing ranges rather than single points, the
single line type tick marks for scatterplots are not appropriate for
gatherplots; instead, ticks should communicate the partitioned
segments on the axes.
Without this visual representation, when the user is confronted with a
number, it can be confusing to determine whether adjacent nodes with
different offset has same value or not.

\begin{figure}[htb]
  \centering
  \includegraphics[width=\columnwidth]{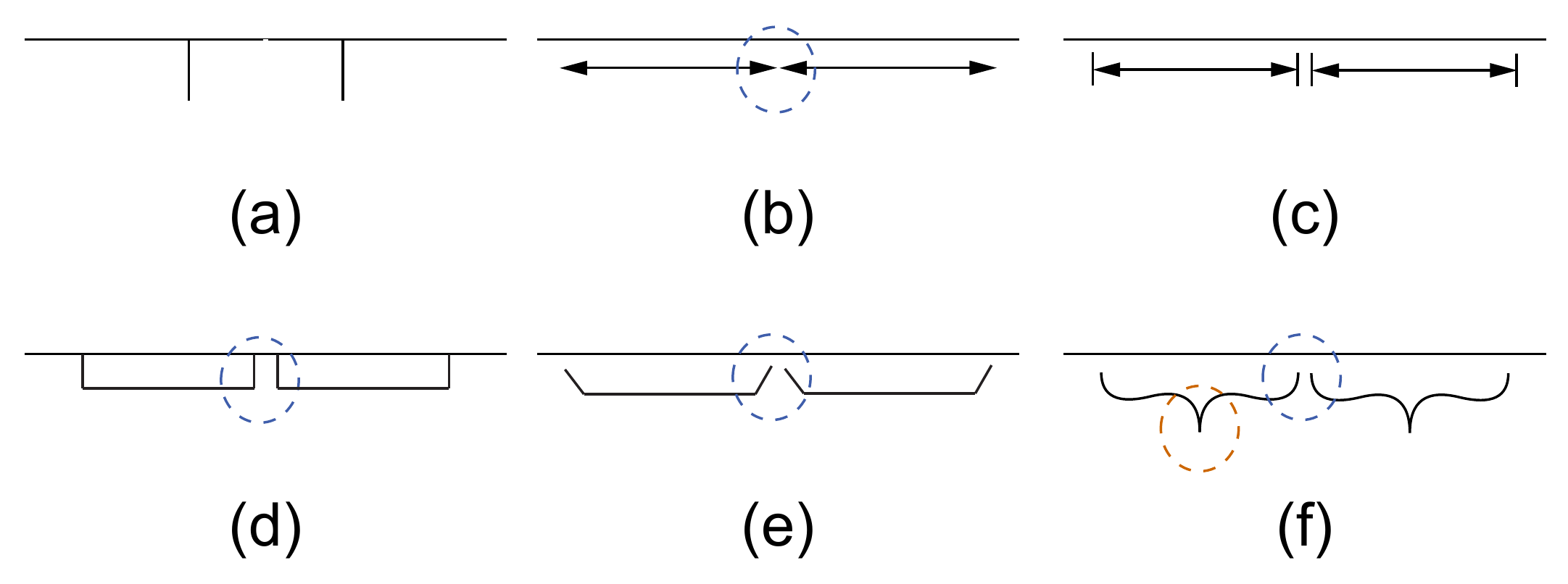}
  \caption{\textbf{Various tick mark types.}
    The blue dotted region represents the area between adjacent tick
    marks.
    (a) is a typical line type tick mark for scatterplots.
    (b) lacks guidelines, which will make anchoring easier.
    (c) creates a packed region between adjacent marks. 
    (d) uses less clutter, but (e) is the least cluttered.
    (f) is the final recommendation, with the data label in the orange
    region.}
  \label{fig:tickMark}
\end{figure}

After considering a few visual design alternatives, we recommend a
bracket type marker for this purpose.
Figure~\ref{fig:tickMark} shows design alternatives of tick mark for
representing ranges.
The bracket is optimal in that it uses minimal ink and creates less
density with adjacent ticks.

\subsection{Interaction}

Gatherplots support the same types of interactions as scatterplots.
However, some additional interaction techniques are required to specifically control the gathering transformation. 

\begin{figure}[htb]
  \centering
  \includegraphics[width=\columnwidth]{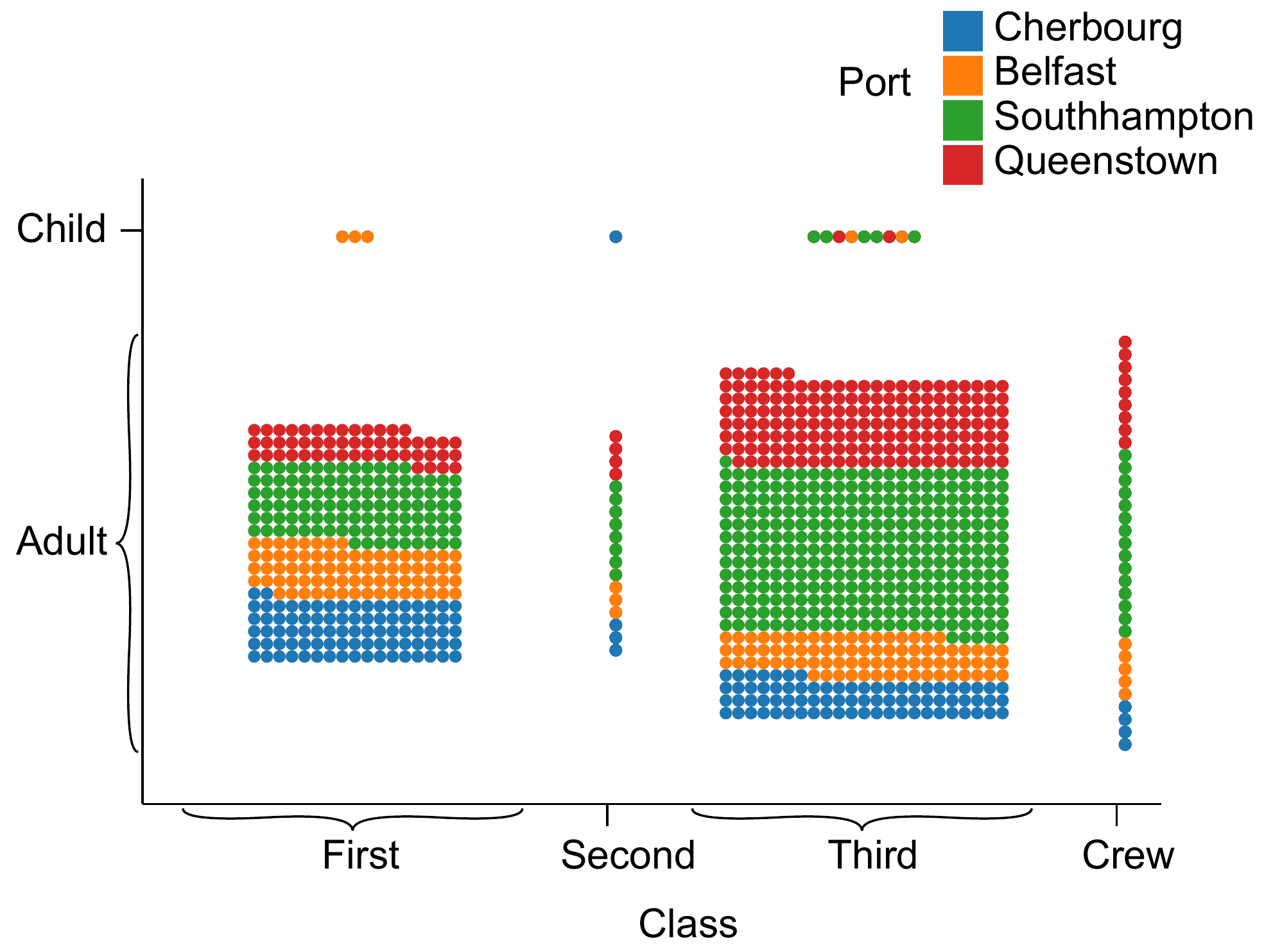}
  \caption{\textbf{Survivors of the Titanic using gatherplots.}
    The X axis is class of passengers, where second class passengers and crew are minimized. 
    The Y axis is age, where the adult value is maximized.
    This view makes it easy to compare first class adults and third class adults. 
    Note that even in the minimized state, we can get an overview about the second class and crew by the color line, which
    communicates the underlying distribution.
    This is due to sorting over the color dimension.}
  \label{fig:axisFolding}
\end{figure}

For example, when exploring multidimensional datasets, it is crucial to have a mechanism to filter unwanted data.
To support this process in gatherplots, we provide an optional mechanism to go back to the original continuous linear scale function.
We allow each axis tick have an interactive control to be filtered out (minimize) or focused (maximized).
This is called \textit{axis folding}, because it can be illustrated by folding a paper.
When minimized or folded, the visualization space is shrunk by applying linear scales instead of non-linear gather scales.
This results in overplotting, as if a scatterplot was used for that axis.
Maximization simply folds all other values except the value of the interest to assign maximum visual space to that value.
Figure~\ref{fig:axisFolding} shows axis folding applied to second class adult passengers in the Titanic dataset.

%% ---------------------------------------------------------------------
%% IMPLEMENTATION
%% ---------------------------------------------------------------------
\subsection{Implementation}

We have implemented a web-based demonstration of gatherplots using D3.js\footnote{\url{http://www.d3js.org}} and
Angular.\footnote{\url{http://www.angularjs.org}}
The prototype allows users to load various datasets into a gatherplot.
The visualization can be compared to scatterplots and jittered scatterplots with a single click.
In the top right area, an interactive guide is provided where users can follow step-by-step instructions in a guided tour of gatherplots.

%% ---------------------------------------------------------------------
%% EVALUATION
%% ---------------------------------------------------------------------
\section{Gatherplots: Evaluation}

This study was designed to demonstrate the effectiveness of gatherplots, in particular its different layout modes with categorical vs.\ categorical variables.
Crowdsourcing platforms have been widely used and have shown to be reliable platforms for evaluation studies~\cite{Paolacci2010, Willett2013}.
Therefore, we conducted our experiment on Amazon Mechanical Turk.\footnote{\url{https://www.mturk.com}}
This also gave us the opportunity to study the utility of the technique for the general population, who do not have specific statistical training.

\subsection{Experimental Design}

We selected jittered scatterplots as the baseline condition, as this technique is a widely accepted standard technique maintaining consistency with scatterplots.
We also wanted to measure the efficiency of different modes of gatherplots.
Therefore, we designed the experiment to have four conditions: scatterplots with jittering (jitter), gatherplots with absolute mode (absolute), gatherplots with normalized mode (normalized), and gatherplots with a toggle to switch between absolute and normalized mode (both).
We adopted a between-subjects design to eliminate learning effects from experiencing other modes.

\subsection{Participants}

A total of 240 participants (103 female) completed our survey.
Because some questions asked about concepts of absolute numbers and probability, we limited demographic to the United States to remove the influence of language.
To ensure the quality of the workers, the qualification of workers were the approval rate of more than 0.95 with number of hits approved to be more than 1,000.
Only three of 240 participants did not use English as their first language.
119 people had more than bachelor's degree, with 42 people having a high school degree.
We filtered random clickers by removing any trials where the completion time was shorter than a reasonable time (5 seconds). 
This yielded a total of 211 participants.

\subsection{Task}

As scatterplots can support many types of tasks, it is difficult to come up with a single representative task.
In the end, we selected retrieving a value as a low-level task, and comparing and ranking as a high-level task.
For the comparing and ranking task, two different types of questions were asked: the tasks to consider absolute values such as frequency and tasks that consider relative values such as percentage.
Therefore, for one visualization, 5 different questions were generated.
For gatherplots, our interest is in the difference between task considering absolute values and relative values.
The five tasks are as follows:

\begin{itemize}
\item\textbf{T1:} retrieve value considering one subgroup.
\item\textbf{T2:} comparing absolute size of subgroup between groups.
\item\textbf{T3:} ranking absolute size of subgroup between groups.
\item\textbf{T4:} comparing relative size of subgroup between groups.
\item\textbf{T5:} ranking relative size of subgroup between groups.
\end{itemize}

To reduce the chance of one chart being optimal by luck for a specific task, two charts of same problem structure were provided.
This yielded a total of 10 questions for each participant.
Each question was followed by the question asking confidence of estimation with a 7-point Likert scale, and the time spent for each question was measured.

\subsection{Dataset}
 
We used a dataset about the Survivors of Titanic.
Each of the 2,201 survivors had four dimensions, which were all categorical variables: class (4 levels), sex (2 levels), port of entry (4 levels), survival status (2 levels).
The five tasks above were asked for two views with different dimensions.
One view visualized class on X-axis, sex on Y-axis, and survival using color.
The second view visualized survived on the X-axis, class on the Y-axis, and port of entry using color.

\subsection{Hypotheses}

We believe that different types of tasks will favor from different type of layouts.
Therefore, our hypotheses are as follows:

\begin{itemize}
\item[H1] For retrieving value considering one subgroup (T1), both absolute and normalized modes will yield better accuracy than jitter mode.

\item[H2] For tasks considering absolute values (T2 and T3), absolute mode will yield the best accuracy over other modes.
  
\item[H3] For tasks considering relative values (T4 and T5), normalized mode will yield the best accuracy over other modes.
  
\end{itemize}

\subsection{Results}

The results were analyzed with respect to the accuracy (correct or incorrect), time spent, and confidence of estimation.
Based on our hypotheses, we analyzed the different modes of layout for each type of question: retrieve value, absolute value task, and relative value task.

\begin{figure*}[htb]
    \centering
    \subfloat{
        \includegraphics[width=0.3\linewidth]{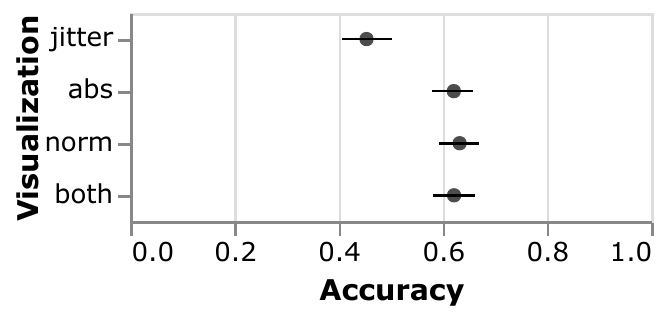}
    }
    \subfloat{
        \includegraphics[width=0.3\linewidth]{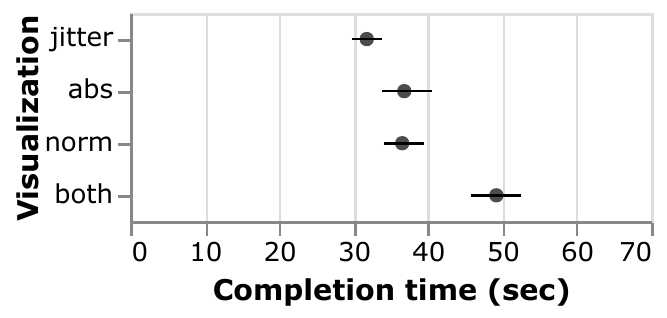}
    }
    \subfloat{
        \includegraphics[width=0.3\linewidth]{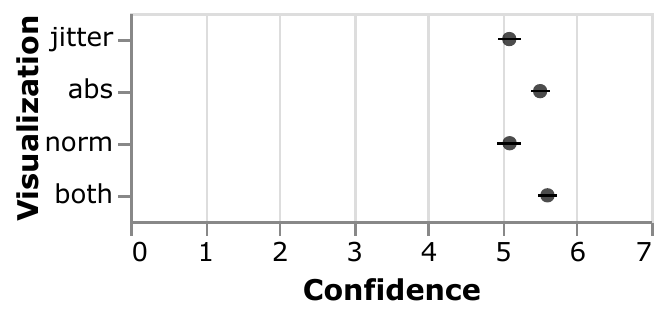}
    }
    \caption{\textbf{Performance results.}
    Effect of visualization on accuracy, completion time, and perceived confidence.
    Error bars show 95\% confidence intervals calculated using bootstrapping ($N=1,000$ repetitions).}
  \label{fig:results-boot}
\end{figure*}

\begin{figure*}[htb]
    \centering
    \subfloat{
        \includegraphics[width=0.3\linewidth]{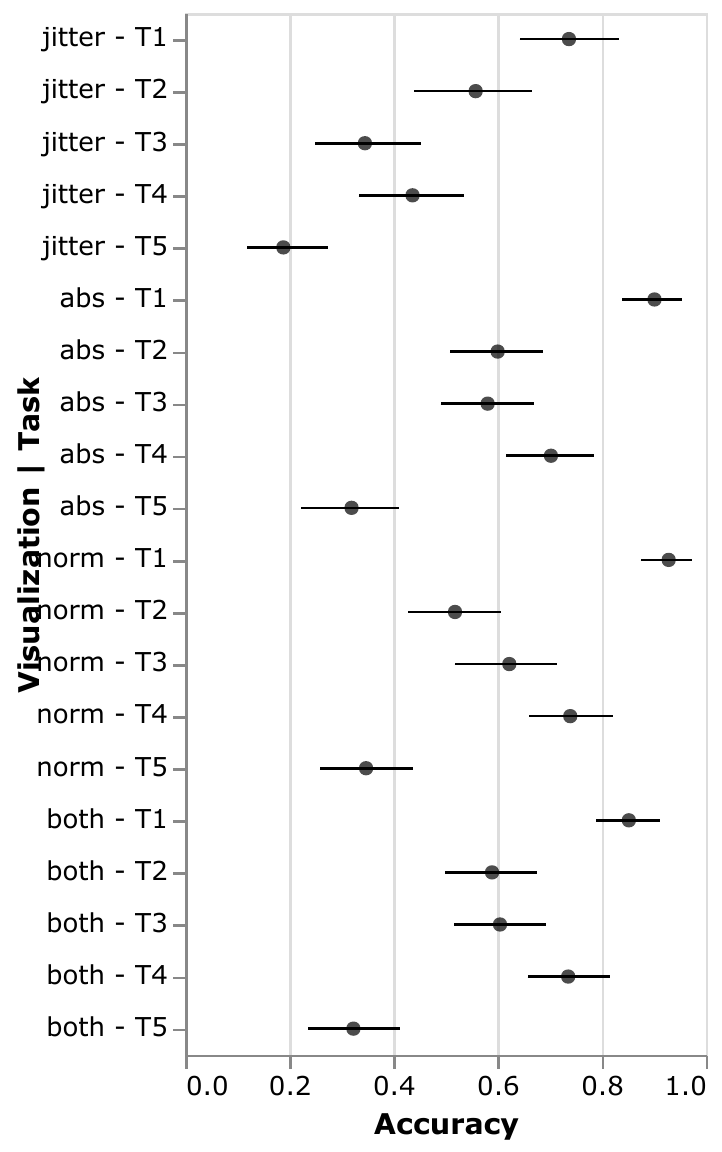}
    }
    \subfloat{
        \includegraphics[width=0.3\linewidth]{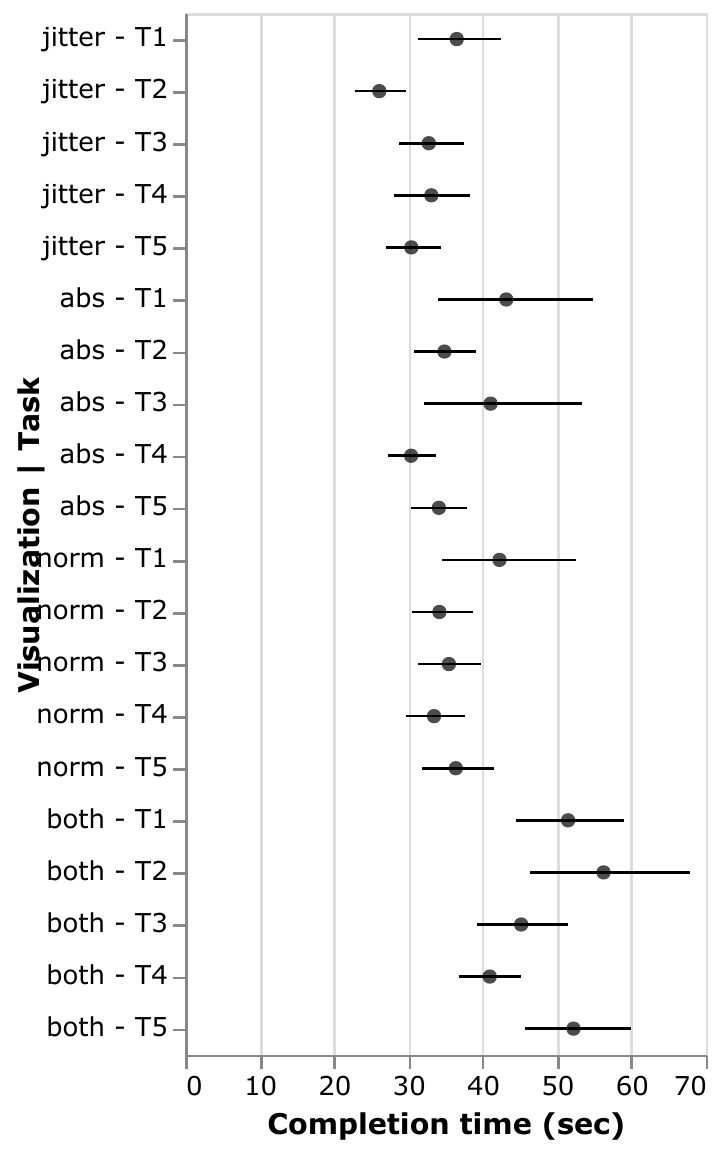}
    }
    \subfloat{
        \includegraphics[width=0.3\linewidth]{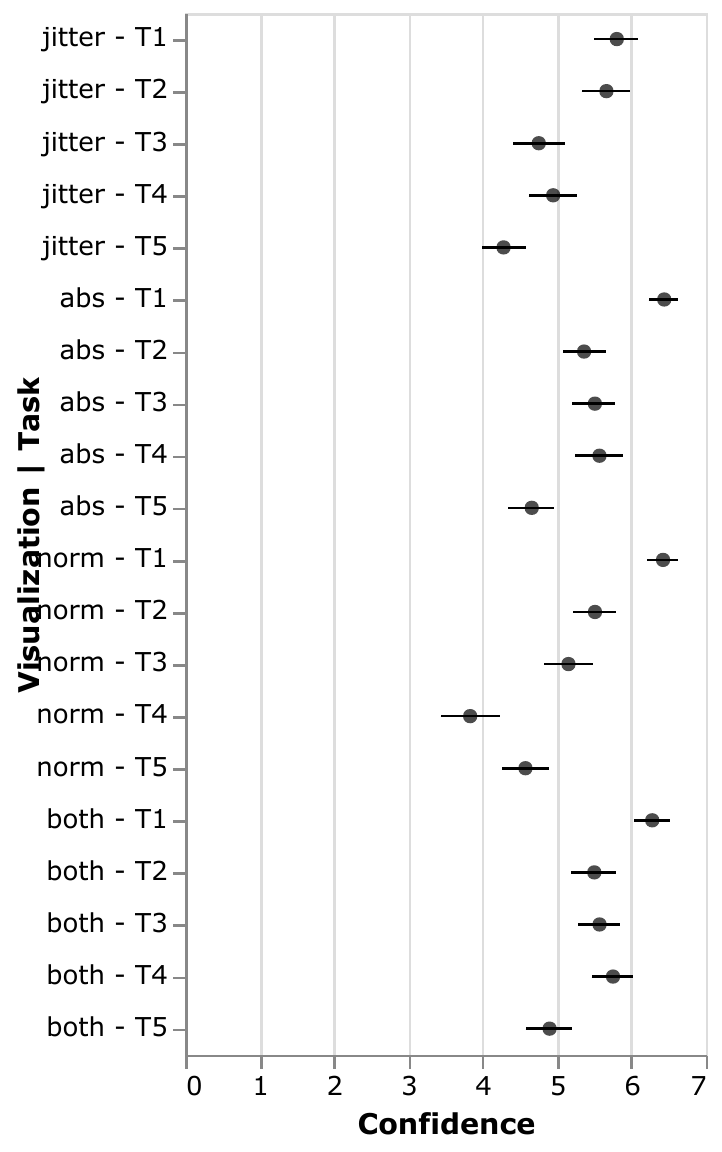}
    }
    \caption{\textbf{Detailed performance results.}
    Effect of visualization and task (T1--T5) on accuracy, completion time, and perceived confidence.
    Error bars show 95\% confidence intervals calculated using bootstrapping ($N=1,000$ repetitions).}
  \label{fig:results-detail}
\end{figure*}

\subsubsection{Accuracy}

\begin{figure*}[htb]
  \centering
  \includegraphics[width=\textwidth]{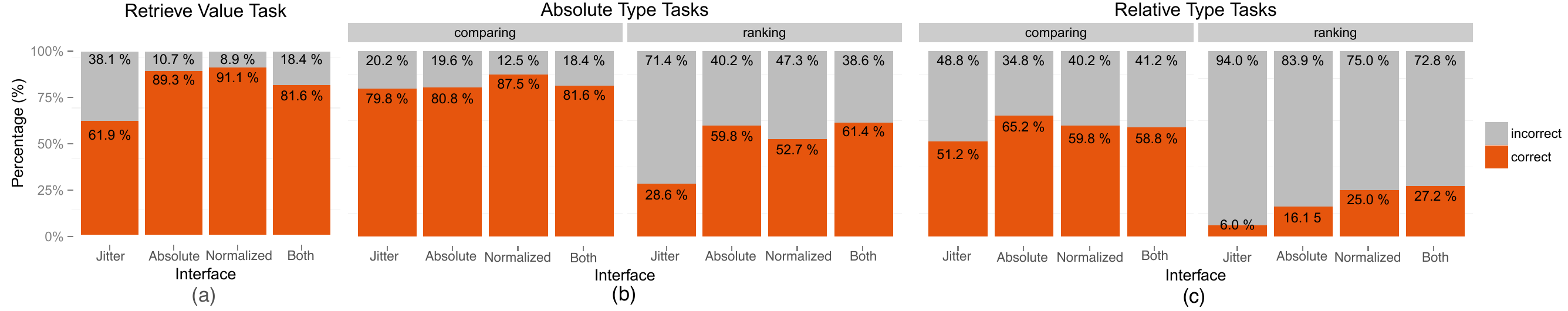}
  \caption{\textbf{Results from our crowdsourced user study.}
    (a) Correctness for retrieving value.
    (b) Correctness for absolute comparison and ranking.
    (c) Correctness for relative/normalized comparison and ranking.}
  \label{fig:correct_type_task}
\end{figure*}

Performance results are summarized in Figure~\ref{fig:results-boot}, with detailed results by visualization type and task in Figure~\ref{fig:results-detail}.
The number and percentage of participants who answered correct and incorrect answers are shown in Figure~\ref{fig:correct_type_task}.
In total, we recruited 42 participants for jitter, 56 participants for absolute, 56 participants for normalized, and 57 participants for interactive mode.

As the measure for each question was either correct or incorrect, we employed logistic regression to analyze the data.
For the retrieving-value task (T1), both the absolute mode and normalized mode had significant main effects (Wald Chi-Square = $18.58$, $p < 0.01$, Wald Chi-Square = 21.05, $p < 0.01$, respectively) with a significant interaction effect (Wald Chi-Square = 19.53, $p = 0.03$) (H1 confirmed).
For absolute-value tasks (T2 and T3), both the absolute mode and normalized mode had significant main effects (Wald Chi-Square = 10.35, $p < 0.01$, Wald Chi-Square = 10.35, $p < 0.01$, respectively) with a significant interaction effect (Wald Chi-Square = 4.31, $p = 0.03$) (H2 confirmed).
For relative-value tasks (T4 and T5), only the normalized mode had a significant effect (Wald Chi-Square= 5.10, $p = 0.02$) (H3 confirmed).

\subsubsection{Completion Time}

We compared the completion time (in seconds) for each task using a mixed-model ANOVA with repeated measures.
For the retrieve-value task, on average, the completion time (sec) for each interface was for jitter 44.26, absolute 56.84, normalized 52.45, and both 56.57.
There was no significant difference between interfaces ($p > 0.05$).

For the absolute-value task (T2 and T3), on average, the completion time (sec) for each interface was for jitter 30.74, absolute 32.3, normalized 33.6, and both 47.91.
The interface had a significant main effect ($F(3, 207) = 11.5, p < 0.01$).

For relative-value task (T4 and T5), on average, the completion time for each interface was for jitter 26.6, absolute 31.12, normalized 31.38, and both 46.78.
The interface had a significant main effect ($F(3, 207) = 10.12, p < 0.01$).
However, when we conducted pairwise comparisons with adjusted $p$-values using simulation, the only significant difference in time spent was when using the both interface, which took longer ($p < 0.01$ for all comparisons).

\subsubsection{Confidence}

The participants self-reported level of confidence was reported using a 7-point Likert-scale rating.
For the value-retrieving task (T1), a Kruskal-Wallis non-parametric test revealed that the type of interface had significant impact on the confidence level ($\chi^2(3) = 74.57 p < 0.01$).
The mean rating for each interface was for jitter 4.8, absolute 6.3, normalized 6.0, and both 6.25.
A post-hoc Pairwise Wilcoxon Rank Sum test was employed with Bonferroni correction to adjust for multiple comparisons.
The jitter interface was significantly lower than the other three modes ($ p < 0.01$ for all cases).
There was no difference between absolute, normalized, and both interfaces.

For absolute-value tasks (T2 and T3), a Kruskal-Wallis non-parametric test revealed that the type of interface had significant impact on the confidence level ($\chi^2(3) = 18.32, p < 0.01$).
The mean rating for each interface was jitter 5.4, absolute 5.7, normalized 5.0, and both 5.8.
A post-hoc Pairwise Wilcoxon Rank Sum test was employed with Bonferroni correction to adjust for multiple comparisons.
The interface with both modes was significantly higher than normalized and jitter mode ($p < 0.01$ for both); however, there was no difference with the absolute mode.
The interface with absolute mode was significantly higher than normalized and jitter mode ($p < 0.01$).

For relative-value tasks (T4 and T5), a Kruskal-Wallis non-parametric test revealed that the type of interface did not have significant impact on the relative tasks ($\chi^2(3) = 4.1, p = 0.2$).
The mean rating was jitter 4.7, absolute 4.9, relative 4.9, and both 4.8.

One possibility for explaining this result is that the relative task is more difficult than the other tasks.
The low correct percentage of questions are also shown in Figure~\ref{fig:correct_type_task}.
To see that, we tested the confidence level between task types.
A Kruskal-Wallis non-parametric test revealed that the type of task had significant impact on the confidence level ($\chi^2(2) = 148.1, p < 0.01$).
The mean rating for retrieving value 5.9, absolute 5.5, and normalized 4.8.
The post-hoc Pairwise Wilcoxon Rank Sum test was employed with Bonferroni correction to adjust for multiple comparisons, and showed that all three task types were significantly different ($p < 0.01$ for all cases).

%% ---------------------------------------------------------------------
%% GATHERLENS
%% ---------------------------------------------------------------------
%% ---------------------------------------------------------------------
%% GATHER LENS
%% ---------------------------------------------------------------------
\section{GatherLens: A Gathering Magic Lens}

\begin{figure}[htb]
    \centering
    \includegraphics[width=\linewidth]{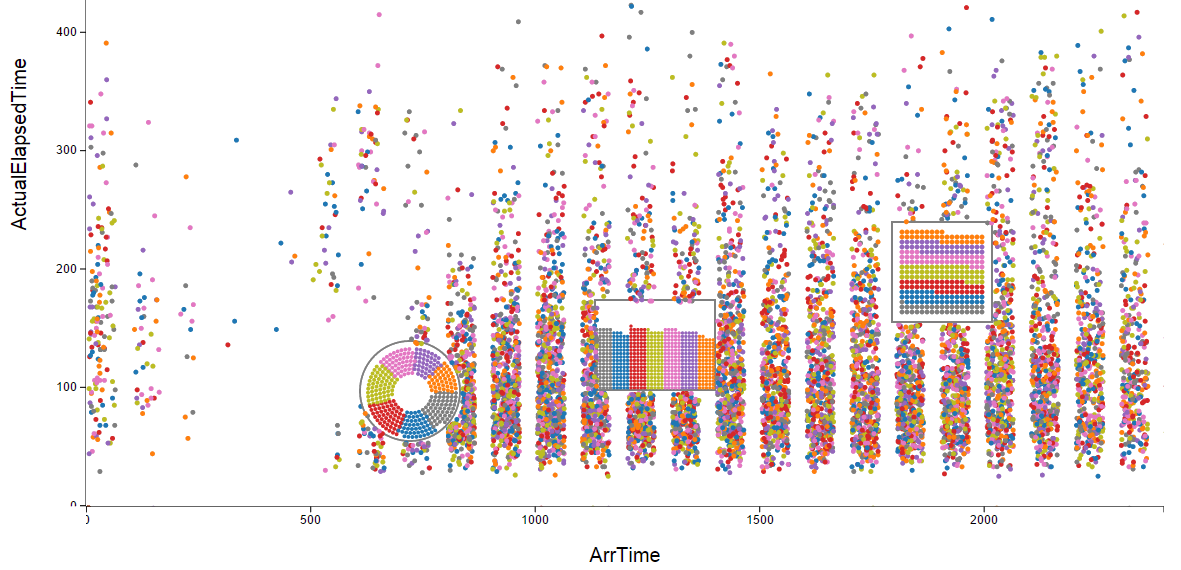}
    \caption{Three examples of a GatherLens in a scatterplot of airline performance data containing 3,000 points.
    From left: pie lens, histogram lens, and standard lens.
    Data point color shows day of the week.}
    \label{fig:gatherlens}
\end{figure}

Scatterplots have a familiar layout and an intuitive continuous scaling in the Cartesian space defined by the graphical axes, whereas gatherplots introduce discontinuities that may make them more difficult to understand.
However, gathering does not necessarily have to be applied globally.
Instead, we here propose a local application of gathering in a Magic Lens technique~\cite{Bier1993} that we call \textit{GatherLens} (Figure~\ref{fig:gatherlens}).

A Magic Lens is a user-controlled interaction tool that changes the visual representation of the underlying graphical object it overlays~\cite{Bier1993}.
The GatherLens is accordingly a Magic Lens that applies local gathering in a scatterplot to the data points that it overlaps.
This gives the user the ability to selectively manage overplotting in specific areas in a scatterplot without changing its overall visual representation.

Lens geometry gives us additional options for layout of the stacked groups in the lens (each is visible in Figure~\ref{fig:gatherlens}):

\begin{itemize}

\item\textbf{Standard lens:} A rectangular lens that applies standard gathering to the contained points.
  
\item\textbf{Histogram lens:} Here, the stacked groups are arranged and aligned so that they resemble a histogram.
  
\item\textbf{Pie lens:} Similarly, the group layout here is radial and centered, yielding a pie or donut layout.
  
\end{itemize}

%% ---------------------------------------------------------------------
%% DISCUSSION
%% ---------------------------------------------------------------------
\section{Discussion}
\label{sec:disc}

Summarizing our results from the user study confirms that gatherplots outperform jittering for both of the object metrics---accuracy and completion time (albeit only marginally for the latter)---as well as for the subjective confidence measure.
While the dataset in the study was only moderately sized, there was still sufficient overplotting to make normal scatterplots useless, causing us to use jittering as the baseline.

Compared to past work, e.g.\ splatterplots~\cite{Mayorga2013} and GPLOMs~\cite{Im2013}, our main contribution is that gatherplots preserve the identity of marks.
Based on the clutter reduction framework proposed by Ellis and Dix~\cite{Ellis2007}, gatherplots are distortion-based methods, while KDE, histograms, and violin plots are appearance-based.
In this section, we will discuss trade-offs compared to appearance-based methods, as well as situations when gatherplots are appropriate and not.
We also discuss results from our user study.

\subsection{Visual Scalability}

\begin{figure*}[htb]
  \centering
  \includegraphics[width=\textwidth]{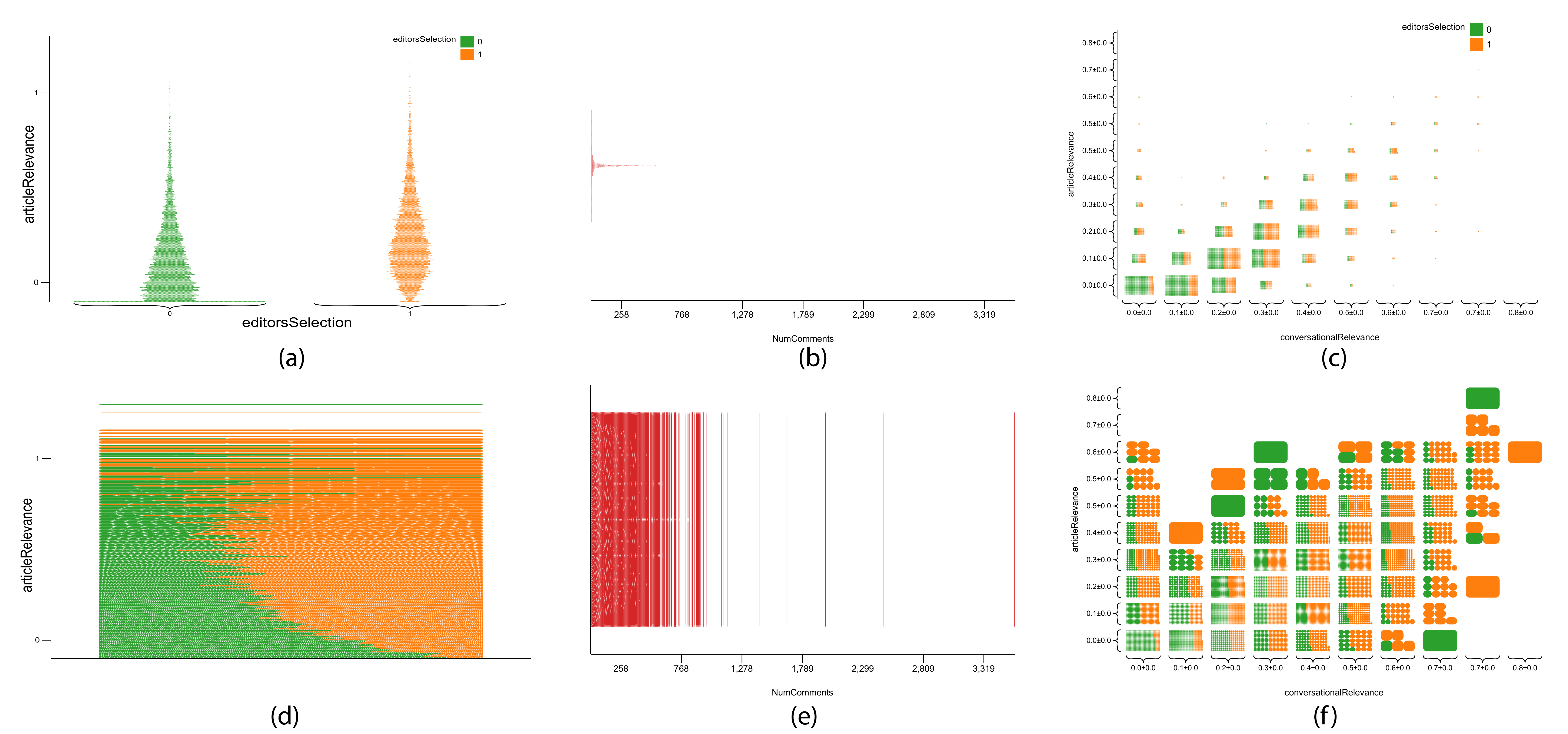}
  \caption{\textbf{Large-scale gatherplots.}
  Three examples of how gatherplots behave in dataset of 30,000 objects.
  (a) resembles a violin plot, while each object maintains its identity.
  (d) shows how relative mode can be used to compare relative distribution.
  In (b) and (c) the outliers become small; however, relative mode in (e) and (f) reveals outliers.}
  \label{fig:scale}
\end{figure*}

As datasets become larger, the visual scalability of a visualization becomes an important issue.
Scatterplots support two main tasks: detecting correlations as well as outliers.
Gatherplots are effective in showing correlations as the dataset grows---as shown in Figure~\ref{fig:scale}(a), (b), and (c)---but this also causes the dot size to shrink, which makes detecting outliers becomes less plausible.
Splatterplots~\cite{Mayorga2013} handle this by using two different visual representation for dense areas and sparse areas, whereas gatherplots have no such mechanism.
In gatherplots, the relative mode enlarges outliers to fill the assigned space.
This makes spotting outliers in gatherplot easier (Figure~\ref{fig:scale}(e) and (f)), as well as to compare distributions (Figure~\ref{fig:scale}(d)).
As datasets grow in size, individual object identification becomes less relevant, and gatherplots begin to resemble histograms or violin plots. 

Even though gatherplots can theoretically be extended to virtually unlimited datasets, the practical bottleneck lies in calculating layouts for individual objects, which requires heavy computation compared to appearance-based methods.
Also, the memory required to handle individual object independently is another bottleneck.  
These computations are hard to justify for a large dataset, because the amount of output information is nevertheless same.
In this sense, gatherplots are not scalable to large datasets. 
However, according to Shneiderman's visual information seeking mantra---\textit{overview first, zoom and filter}~\cite{Shneiderman1996}---even for large datasets, zooming can make the application of gatherplots desirable.  

\subsection{Named vs.\ Anonymous Objects} 

In some multidimensional datasets, data points have names, whereas in others they do not.
For example, in datasets containing entities such as cars or digital cameras, it is a common task for users to search cases that suit their needs.
For this case, maintaining the identity of individual data points is important because it enables brushing and linking more easier than for aggregated forms such as histograms. 

There are other datasets, such as the famous iris flower dataset, where individual data points do not have names.
For such datasets, the benefit of maintaining object identity can instead be explained using frequency grids.
According to Cosmides and Toody~\cite{Cosmides1996}, the concept of relative percentage is a new concept in human evolution, and explicitly showing distribution using discrete countable objects makes assessing the relative percentage easier and more comfortable.

\subsection{Visualizing Normalized Data}

According to Im et al.~\cite{Im2013}, one tradeoff in designing GPLOMs was whether axes of the same variable should be scaled to the same range or not.
If scaled to the same range, it would be easy to compare to adjacent charts, but results in large vacant spaces for sparse areas.
This happens when there are severe imbalances in data distribution.
Both options are valid depending on the task at hand, but it is difficult to represent it visually in histograms so that users can see it.

In gatherplots, these two options are supported as absolute vs.\ normalized modes. 
Because we show the individual objects as separate visual marks, it is feasible to deliver this information more explicitly.
If the size of all data points are the same, users will understand that they all use the same scale, while rendering points with different sizes suggests that they are normalized.

\subsection{Limitations of Evaluation Study}

Although scatterplots support several types of tasks such as detecting correlation, clusters, or outliers, in our experiment we decided to test a particular case with categorical data, which has distinctive views compared to conventional scatterplots.
Even if this is a narrow case, the purpose of our study was to show the effectiveness of different layout modes in a quantitative way.
The results indicated that the users could understand the visualization and accomplish the tasks that should be supported.
However, we also observed that the difficulty level was different for each task type.
In general, ranking tasks were more difficult than comparison tasks, and questions asking about relative values were more difficult than those about absolute values.
Therefore, maintaining similar difficulty level among tasks should also be considered while designing future comparative evaluations such as this.

In our study, we selected scatterplots with jittering as the baseline for comparison because (1) it extends scatterplots to manage overplotting, (2) it maintains individual objects, and (3) it is a well-known technique.
However, for future studies it would be also desirable to compare the performance with a purpose-specific technique, such as basic scatterplots, histograms, or hieraxes.

%% ---------------------------------------------------------------------
%% CONCLUSION
%% ---------------------------------------------------------------------
\section{Conclusion and Future Work}

We have proposed the concept of the gather transformation, which
enables space-filling layout without overdrawing while maintaining
object constancy.
We then applied this transformation to scatterplots, resulting in
gatherplots, a generalization of scatterplots, which enable overview
without clutter.
While gatherplots are optimal for categorical variables, it can also
be used to ameliorate overplotting caused by continuous ordinal
variables.
We discussed several aspects of gatherplots including layout,
coloring, tick format, and matrix formations.
We also evaluated the technique with a crowdsourced user study showing
that gatherplots are generally more effective than jittering, and
absolute and relative mode serve specific types of tasks better.
We also applied the gathering transformation to a Magic Lens
interaction for local control; this lens has three different layout
modes.

We believe that gathering is a general framework that captures the
transition between overlapping and space-filling visualizations while
maintaining object identities.
In the future, we plan on studying the application of this framework
to other visual representations.
For example, overplotting is a common problem when visualizing
categorical variables in a parallel coordinates plot.
Parallel sets aggregate elements for the same value of a categorical
variable into blocks, but loses the identity of objects.
By applying the gathering framework, parallel sets can be
reconstructed to render individual lines instead of block lines, which
would enable combining both categorical and continuous variables.
Furthermore, we also want to study additional gathering-based
interaction techniques beyond the GatherLens proposed here.

%% ---------------------------------------------------------------------
%% ACKNOWLEDGMENTS
%% ---------------------------------------------------------------------
\section*{Acknowledgments}

We would like to thank S.\ Karthik Badam, Jungu Choi, and Senthil K.\ Chandrasegaran for helpful discussions and Ji Soo Yi for advice on our evaluation.
Also, we would like to acknowledge all the help we received from communicating with a number of visualization creators, some of whose works are cited in the paper.

%% ---------------------------------------------------------------------
%% Bibliography
%% ---------------------------------------------------------------------

\bibliographystyle{plainnat}
\bibliography{gatherplots}

\end{document}